\documentclass[12pt,preprint]{aastex}

\def\lsd{L_{\rm SD}}
\def\Pdot{\dot P}
\def\psr{PSR B1257+12}

\def\spit{{\it Spitzer}}

\def\70um{70~\micron}
\def\160um{160~\micron}
\def\24um{24~\micron}
\def\um{\micron}
\def\ld{L_{\rm dust}/L_{\star}}
\def\ldsd{L_{\rm dust}/L_{\rm SD}}

\def\Lstar{L_{\star}}
\def\MEarth{M_\oplus}
\def\REarth{R_\oplus}

\def\LSun{L_\odot}

\def\gapp{\lower 3pt\hbox{${\buildrel > \over \sim}$}\ }
\def\lapp{\lower 3pt\hbox{${\buildrel < \over \sim}$}\ }
\def\proptosim{\lower 3pt\hbox{${\buildrel \propto \over \sim}$}\ }

\begin{document}

\title{\spit/MIPS Limits on Asteroidal Dust in the \\ Pulsar Planetary System PSR B1257+12}

\author{
G. Bryden$^{1}$, C. A. Beichman$^{2}$, 
G. H. Rieke$^{3}$, \\
J. A. Stansberry$^{3}$, 
K. R. Stapelfeldt$^{1}$, 
D. E. Trilling$^{3}$, 
N. J. Turner$^{1}$, \& 
A. Wolszczan$^{4}$
\vskip 0.05in
\affil{1) Jet Propulsion Lab, 4800 Oak Grove Dr, Pasadena, CA 91109} 
\affil{2) Michelson Science Center, California Institute of Technology, 
  Pasadena, CA 91125}  
\affil{3) Steward Observatory, University of Arizona, 933 North
Cherry Ave, Tucson, AZ 85721}
\affil{4) Department of Astronomy and Astrophysics, Pennsylvania State
University, University Park, PA 16802}
\email{bryden@jpl.nasa.gov}}

\shorttitle{Dust Limits for PSR B1257+12}
\shortauthors{Bryden et al.}

\begin{abstract}
With the MIPS camera on \spit, we have searched for far-infrared
emission from dust in the planetary system orbiting pulsar \psr.
With accuracies of 0.05 mJy at \24um and 1.5 mJy at \70um, 
photometric measurements find no evidence for emission at these wavelengths. 
These observations place new upper limits on the luminosity
of dust with temperatures between 20 and 1000 K.
They are particularly sensitive to dust
temperatures of 100-200 K, for which they limit the
dust luminosity to below $3 \times 10^{-5}$ of the pulsar's spin-down
luminosity, three orders of magnitude better than previous limits.
Despite these improved constraints on dust emission, an asteroid belt
similar to the Solar System's cannot be ruled out.
\end{abstract}

\section{Introduction}

Before the discovery of extrasolar planets around main-sequence stars
\citep{mayor95,marcy95}, pulsar timing measurements provided the first
evidence for an extrasolar planetary system \citep{wolszczan92}.
The initial discovery of two planets was later expanded to three:
$0.02 \pm 0.002 \ \MEarth$ at 0.19 AU,
$4.3 \pm 0.2 \ \MEarth$ at 0.36 AU, and
$3.9 \pm 0.2 \ \MEarth$ at 0.46 AU
\citep{wolszczan94, wolszczan00, konacki03a}.
Orbital analysis of the pulsar timing measurements reveals a coplanar
system with the outer planets in 3:2 orbital resonance \citep{konacki03a}, 
strongly suggesting that their formation mechanism must
involve a preplanetary disk of material circling the neutron star.
A variety of theories have been proposed for the formation of such a disk
in this system \citep[see][]{miller01} -
from the remains of a merger event \citep{podsiadlowski91},
from the disruption or ablation of a stellar companion
\citep{stevens92, tavani92}, or
from the fall back of supernova ejecta \citep{Lin91}.
The presence of an early disk is further
motivated by theories for millisecond pulsar formation, which generally
use an accretion disk to spin up the pulsar \citep{michel85}.

In its presumed disk origin, the pulsar system's history is thought to
be similar to our own Solar System and its protostellar nebula.
Evidence for such disks around hydrogen-burning stars is clear.
Massive protoplanetary disks are commonly found in young star forming
regions, both inferred from spectral energy distributions
\citep[][]{beckwith90}  
and seen directly in silhouette \citep{mccaughrean96}.
The older remnants of these disks were first discovered around main
sequence stars by IRAS \citep{aumann84}, with many debris
disks later identified by ISO \citep[][]{habing01} and now with
\spit\ \citep[][]{rieke05}.  
Dusty disks are frequently found around main sequence stars  
over a range of spectral types and ages, yet
no disk emission has ever been detected from around a pulsar.

Many attempts have been made to survey nearby pulsars for dust
emission.
The broadest of these surveys, an examination of the positions of 478
pulsars in the IRAS Point Source Catalog, failed to identify any
excess beyond that expected from coincidental alignment \citep{vanburen93}.
Many smaller but more sensitive surveys of pulsars have been made,
both in the infrared \citep{foster96, koch02, lazio04} 
and at sub-mm wavelengths \citep{phillips94, greaves00, lohmer04}.
As the host to a planetary system, \psr\ has been targeted in
particular, both within surveys and by other specific observations.
A range of wavelengths have been considered
from visible \citep{Abazajian05}
to near-IR \citep[][]{cutri03},
mid-IR \citep[][]{foster96},
far-IR \citep[][]{moshir90},
and sub-mm \citep{phillips94, greaves00, lohmer04}.
(A summary of these \psr\ observations is shown in Figure~\ref{obs} below.)
In each case, only upper limits for dust emission were obtained.

Nevertheless, these searches have been limited by their sensitivity,
particularly at far-IR wavelengths.
Very high upper limits for the dust mass ($\sim$100 $\MEarth$) are
commonly cited \citep{lohmer04,lazio04}.
Pulsars tend to be much farther away than the
main-sequence stars identified as having debris disks
\citep[typically tens of pc for solar-type stars;][]{bryden06a}, 
making detection more difficult. 
Perhaps more importantly, the pulsar's efficiency in heating any dust
that might be present may be much lower than for hydrogen-burning
stars whose radiation peaks in visible light (see \S\ref{dustlim} below).
Given the difficulty in detecting dust around pulsars, the {\it
Spitzer Space Telescope}, with unprecedented sensitivity to 
infrared radiation, is an ideal observatory for continuing the ongoing
search for dust in the \psr\ system.  
Below, we describe such \spit\ observations (\S\ref{obssect}) and then
use them to place stricter limits on the \psr\ dust luminosity
(\S\ref{dustlim}). 

\section{Observations}\label{obssect}

\psr\ was observed with the \spit\ long-wavelength camera, MIPS, 
on 21 Jun 2005 at both 24 and \70um.  
Our overall data analysis is similar to that previously described by
\citet{beichman05mips}, \citet{bryden06a}, and \citet{gautier06}.
At \24um, mosaiced images were created from the raw data using the DAT
software developed by the MIPS instrument team \citep{Gordon05a}.  
Several additional corrections were applied, including the subtraction
of smooth gradients across the field (to remove scattered light
effects parallel to the scan mirror direction)
and the application of a second-order flat, derived from the data
itself (to correct for dark latents, residual jail-bars, and broad
flat-field trends).
At \70um, images were processed beyond the standard DAT software 
to correct for time-dependent transients, corrections
which can greatly improve the sensitivity of the measurements
\citep{Gordon05b}.
For both wavelengths, aperture photometry was performed using
calibration factors, apertures sizes, background annuli, and aperture
corrections as in \citet{beichman05mips}. 
\psr\ was not observed at \160um, due to the
high background noise in MIPS images at that wavelength 
(typically tens of mJy) relative to existing sub-mm limits.

At the observed wavelengths,
we fail to detect significant emission from the pulsar system.
Upper limits are calculated directly from the noise levels measured
within each field.
At \24um, where we have integrated over 5 cycles of 3 sec exposures,
we achieve a (1-$\sigma$) sensitivity of 0.045 mJy.
At \70um, with 5 cycles of 10 sec exposures, our sensitivity is 
1.45 mJy.
This level of accuracy at \70um is better than that typically seen
in similar observations \citep[e.g.][]{beichman05mips}, a reflection of
the fortuitously low background level in this field.
The accuracy of this observation is likely to be limited by confusion
with background extragalactic sources, such that longer integration
time would not significantly improve the results \citep{dole04,bryden06a}.

The 3-$\sigma$ upper limits calculated here are 
shown in Figure~\ref{obs} alongside those from previous investigations.
While $\sim$mJy limits have been obtained at near-IR and sub-mm 
wavelengths, previous observations have been least sensitive at the
far-IR wavelengths most commonly used to detect dust emission around
main-sequence stars.
At \24um, the \spit\ limit is more than three orders of magnitude better
than IRAS, while at \70um there is a factor of $\sim$40 improvement.
The following section translates these limits on flux into
constraints on the dust temperature and luminosity.

\section{Dust Constraints}\label{dustlim} 

Other searches for dust around \psr\ have generally compared their
observational limits with the dust model of \citet{foster96}.
In this model, the dust absorbs and re-radiates some fraction of the
energy that the pulsar is known to be losing as it spins down.
This spin-down luminosity, $\lsd =4 \pi^2 I \Pdot/P^3$, can be easily
calculated from the pulsar's period, $P$, and spin-down rate, $\Pdot$.
The pulsar's moment of inertia, $I$, is assumed to be $10^{45}$ g cm$^2$
\citep{foster96}.
For \psr\ 
\citep[$P = 6.2$ ms, $\Pdot = 1.1\!\times\!10^{-16}$ ms/s;][]{konacki03a}
the spin-down luminosity is $5.2 \ \LSun$.
Following \citet{foster96}, previous work \citep[e.g.][]{greaves00,lohmer04}
typically assumed a dust luminosity of 1\% of $\lsd$,
a plausible upper limit for the fraction of energy a thick dust disk
might intercept.

Rather than assume an ad hoc dust luminosity, we instead choose to
treat $L_{\rm dust}/\lsd$ as the primary unknown quantity whose value we
can place limits on.
In fact, the dust luminosity can be directly constrained from flux
measurements, depending on the temperature of the dust
\citep[e.g.][]{beichman05mips}.
The temperature, however, is very uncertain.
The standard \citet{foster96} dust models assign temperatures of
10-20~K to the dust, but the physical mechanism for converting the
pulsar's rotational energy into thermal 
dust emission is not specified.
The dust temperature in these models is essentially arbitrary.
The ability of the various-sized dust grains to absorb the pulsar's
emitted energy is not considered, nor is the distance of the dust
from that energy source.

Around solar-type stars, which emit radiation at wavelengths readily
absorbed by micron-sized dust grains, dust at 1 AU can reach
temperatures of several hundred K.
Around pulsars with solar-like luminosities spread over a range of
wavelengths, the absorption is presumably less efficient, resulting in
lower temperatures at the same orbital distances. 
Should a belt of material circle outside the orbits of the \psr\
planets, temperatures of $\sim$50-100~K might be expected, depending
on the pulsar's spectral distribution.
The fraction of \psr's luminosity emitted at various wavelengths is 
directly limited by observations.  
Although the pulsar emission model of \citet{malov2003} predicts a
high X-ray flux ($5 \times 10^{-9}$ erg cm$^{-2}$ s$^{-1}$), 
the lack of detection by Chandra sets an upper limit of 
$6 \times 10^{-15}$ erg cm$^{-2}$ s$^{-1}$ \citep{pavlov06}, 
less than $10^{-6}$ of the spin-down luminosity.
Optical non-detections \citep[e.g.][]{Abazajian05}
provide similarly strict limits for visible emission.
Instead, most of the pulsar's spin-down energy is 
emitted as relativistic particles \citep{Gaensler06}.
In this case, the heating of small dust grains is minimal;
each particle impact with a dust particle will eject a 
nuclide from the system without imparting kinetic energy into the
parent dust grain. 
The heating efficiency for the dust in a thin disk may then be very
low, with correspondingly low temperatures.
Rocks that are larger than the stopping depth for the relativistic
particles \citep[$\sim$100 cm;][]{miller01}, however, will absorb
most of the incoming energy and will be much more efficient at 
converting the pulsar's spin-down energy into thermal radiation.

Unlike the highly beamed radio emission, pulsars' particle winds are 
fairly symmetric, as evidenced by the near-spherical shapes of 
pulsar wind nebulae \citep[e.g.][]{kennel84,Gaensler06}.
A key uncertainty, however, is the location where the pulsar's
magnetic dipole radiation transitions into a relativistic particle wind.   
Close to the pulsar, at its light cylinder radius ($c/\Omega$ = 300 km),
the outward flow is thought to be dominated by the Poynting flux 
($E \times B$) \citep{Gaensler06}.  Observations of pulsar nebulae,
however, find that the flux on larger scales is concentrated in a
particle wind.  For the Crab Pulsar's nebula, for example, modeling of
the energy deposition into the wind termination shock requires that a
large fraction ($\gapp$99\%) of the energy have been converted into a
particle wind before reaching a few pc from the pulsar \citep{kennel84, 
begelman92}.  Exactly where or how transition occurs is not known.  
Assuming that the particle wind has fully developed before reaching 
$\sim$AU orbital distances, the pulsar's spin-down energy will
effectively heat all large bodies in the planetary system; 
otherwise only ionized matter would be strongly influenced.

\begin{figure}
\begin{center}
\includegraphics[angle=-90,width=6.5in]{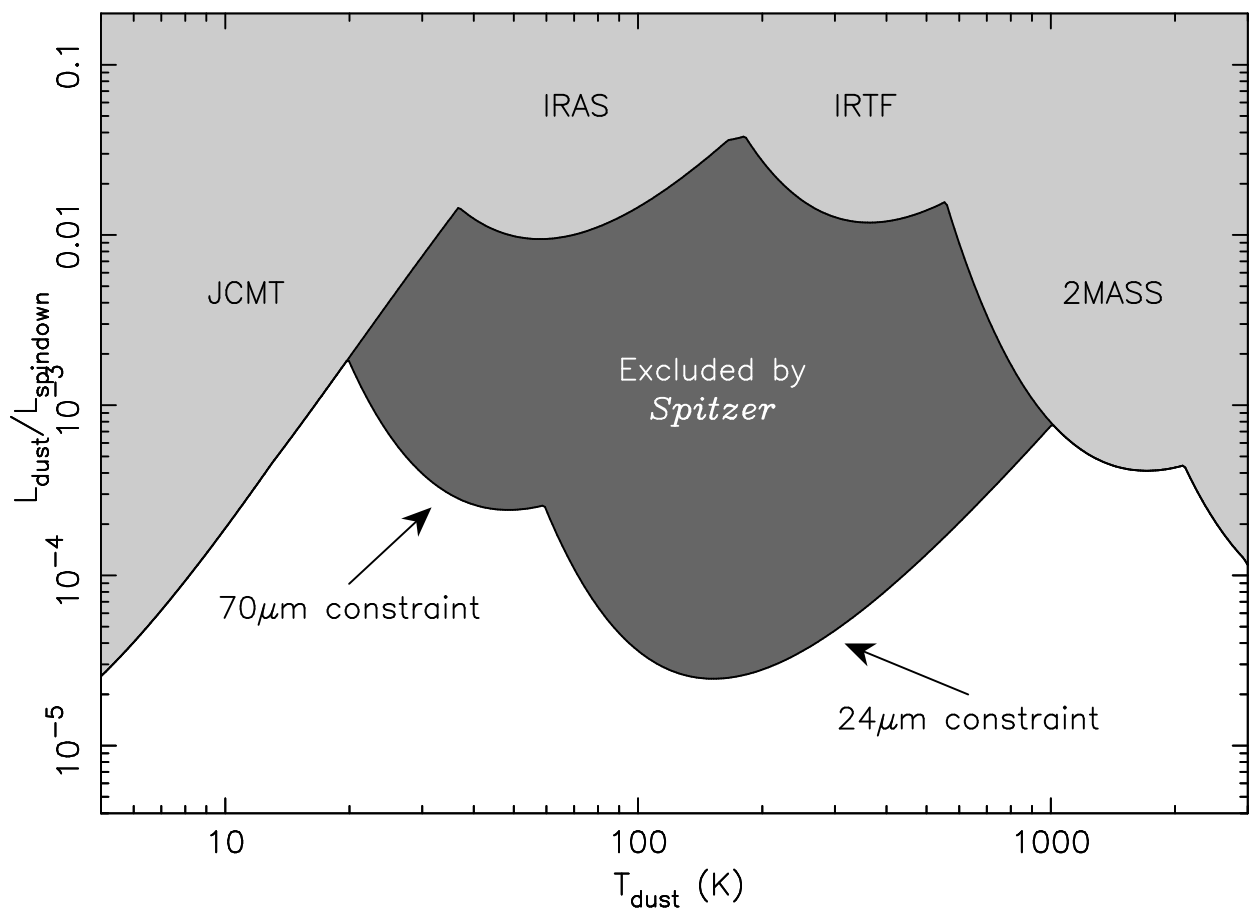} 
\end{center}
\caption{Limits on the temperature and luminosity of dust in the \psr\
system.
The light shaded region is based on previous observations, while the
dark shaded region represents dust parameters ruled out by the 
observations described here.
Dust emissivity is assumed to fall off linearly as the wavelength
increases past 100 \micron.  
To convert fluxes to luminosities, we adopt a pulsar distance of 450 pc
based on the \citet{taylor93} model for the galactic
electron distribution, recently updated by \citet{Cordes02}; 
this distance is uncertain by $\sim$20\%.}
\label{ldls}
\end{figure}

Given the uncertainty in the dust temperature, we leave it as a free
parameter to be constrained by observations.
Figure~\ref{ldls} shows our constraints on $L_{\rm dust}/\lsd$ for a
range of dust temperatures, under the assumption that
the dust can be characterized by a single dominant value.
(In reality some range of temperatures would exist for dust with a
range of grain sizes and orbital locations,
even more so if stochastic heating is responsible for large
temporary increases in dust temperature for a small fraction of the
grains.) 
For each temperature, we calculate the maximum blackbody emission that 
is consistent with the observed 3-$\sigma$ limits.
Each of the curved segments in Figure~\ref{ldls} corresponds
to an individual observation at a specific wavelength, with
the temperatures that an observation is most sensitive to depending 
on that wavelength.
As seen in the figure, the \spit\ observations create more stringent
limits on dust luminosity for dust with temperatures between 20 and 1000 K.
(Above 1000~K, near-IR photometry is more accurate, while below 20~K
sub-mm measurements are more sensitive.)
Better than two orders of magnitude improvement in sensitivity is
achieved over much of this range, 
with limits on $L_{\rm dust}$ as low as $2.5 \times 10^{-5}$ $\lsd$
for $\sim$150~K dust.
For high-emissivity dust orbiting at 2.5 AU,
this corresponds to an upper limit on the emitting area of 
$5 \times 10^{23}$ cm$^2$.

\section{Discussion}

Based on our \spit\ observations, we constrain the luminosity of
dust emission around \psr\ to be $< 10^{-4} \ \lsd$
for temperatures of $\sim$50 to 500~K.
This limit allows us rule out a dense, thick disk that absorbs and
thermalizes a large fraction of the emitted energy -
essentially the same general conclusion as reached by previous authors.
This non-detection of IR excess, however, does not exclude the
presence of an optically thin debris disk orbiting the pulsar.
While the \spit\ observations represent 
a great improvement over previous IR observations,
they still do not have an optimal level of accuracy in terms of 
$L_{\rm dust}/ \lsd$.
Observations of nearby main sequence stars at this sensitivity level
would be able to detect only the brightest debris disks.
For example, only $\sim$2\% of solar-like type stars have luminosities
$> 10^{-4} \ \Lstar$ \citep{bryden06a}.
A much greater fraction of these stars ($\sim$12\%)
have disks within an order of magnitude below this limit,
while the Solar System has $\ld$ even lower
\citep[$\sim$10$^{-7}$ to $10^{-6}$ for the Kuiper Belt;][]{Stern96}.
Even if the dust is efficiently heated by the pulsar,
there may simply be too little for us to detect.

A belt of planetesimals circling at a few AU, similar
to our own asteroid belt, might still be expected.
In other systems, the detection of emitting dust outside of the orbits
of known extrasolar planets \citep{beichman05mips}, similar to the Kuiper Belt
in our Solar System, suggests that an architecture of outer debris belts
encircling inner planets may be a general consequence of planet formation.
\citet{goz05}, meanwhile, have
investigated the long-term stability of test particles
orbiting in the \psr\ system, 
finding that most of the region outside of 1 AU is stable to
perturbations from the three inner planets.
While high energy particles emitted from the pulsar should evaporate
very small bodies on a relatively short time scale, \citet{miller01}
estimate that planetesimals larger than $\sim$km in size can survive
the lifetime of the pulsar.
(The age of the pulsar is
estimated from its spin-down time scale, $P/2\Pdot$, to be $<$Gyr.)
In fact, there is now evidence for an asteroid-like object
orbiting at 2.4 AU with a mass of $4 \times 10^{-4} \ \MEarth$,
about twice the mass of Ceres, the largest asteroid in the Solar System
\citep{wolszczan06}.
Ongoing pulsar timing measurements will probe down to
even smaller masses and can eventually determine whether
many large asteroids are present in this system.

From this pulsar timing, some bounds can be placed on the
total mass contained in larger planetesimals.
Pulsar timing measurements provide direct constraints on the mass
distribution, which is restricted to large asteroid-like mass
concentrations.  For an extended disk, the total mass potentially
contained in asteroids is limited by the requirement that the belt not
grind itself away over lifetime of the system.  
If the age of the pulsar is $\sim$Gyr, the time scale for destructive
collisions between the largest objects must be longer in order for
them to survive.
A disk with tens of Earth masses of planetesimals would break itself
down into smaller objects relatively quickly.
The Solar System's asteroid belt, by comparison, is evolving on
gigayear time scales \citep{dermott02}.
Similarly, the collisional time scale for a disk of
planetesimals a few AU from a pulsar is of order
$\sim R_{\rm pl}\MEarth / \REarth M_{\rm disk}$ Gyr,
where $R_{\rm pl}$ is the typical size of the planetesimals
and $M_{\rm disk}$ is their combined mass
\citep[see e.g.][]{dominik03}, such that a 1-$\MEarth$ disk of
planetesimals would have ground down already.

The mass contained in smaller bodies is less clear.
Given the possibility of weak coupling between dust and the energy
emitted by the pulsar, the observational constraints on dust
luminosity do not translate well into upper limits on dust mass in the
system.  
Rapid dust removal, however, suggests that the dust mass not be 
arbitrarily large.
If the disk is optically thin toward the pulsar wind, 
dust particles will be ablated by relativistic particle impacts on
time scales of $<$~1~yr \citep{miller01},
i.e.\ even faster than the removal of Solar System dust by
Poynting-Robertson drag \citep[$\sim$10$^3$ yr for \um-sized
grains;][]{gustafson94}. 
As in our own system, dust could be continually replenished by
a collisional cascade from larger rocks, but the fast removal 
process would limit the amount of dust to a level below that from a
similar cascade around a main-sequence star.

In order to assess the effect of this ablation on the infrared emission
from a pulsar encircling debris disk,
we consider a collisional cascade model in which
large planetesimals are continuously shattered to produce smaller and
smaller objects.
The equilibrium slope of the resultant distribution
is $dn/da \propto a^{-3.5}$ \citep{dohnanyi68},
such that the smallest grains dominate the overall surface area
while the largest bodies comprise the bulk of the system's mass.
On top of this standard collisional cascade we have included the important
new effect in a pulsar environment - particle ablation by the 
pulsar wind.
Otherwise the simulations we present here are relatively simple
compared to detailed asteroid belt models \citep[e.g.][]{durda97}.
In particular, we assume that
1) the binding strength is independent of object size,
2) impactors can catastrophically disrupt objects up to ten
times their size, 
and
3) the distribution of debris created in such an impact goes as 
$dn/da \propto a^{-3.5}$.
While these simplifications neglect some of the physics
(for a full discussion and detailed models see \citet{davis02}),
they allow us to calculate illustrative models of the 
dust surrounding a pulsar.

\begin{figure}
\begin{center}
\includegraphics[angle=-90,width=6.5in]{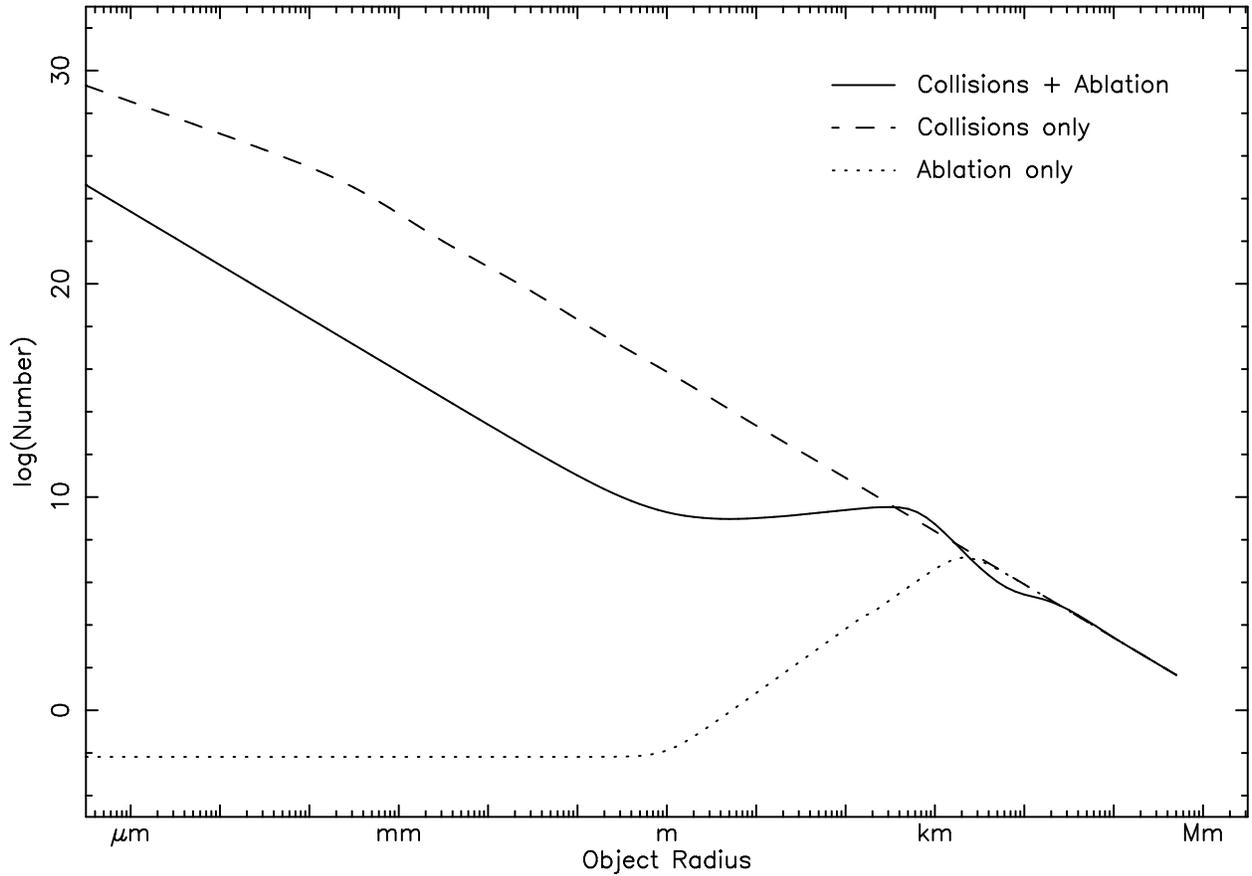} 
\end{center}
\caption{Distribution of particle sizes in a system with 
strong ablation by a relativistic particle wind.
Sizes ranging from \micron-sized dust up to 500 km radius
planetesimals are considered.
The number of particles is given per logarithmic size bin.
Below $\sim$1 km in radius, objects are ablated by the pulsar wind faster 
than they are replenished by collisions of larger sized objects. 
The distribution of particle sizes produced by a standard
collisional cascade is shown for comparison (dashed line),
as is the case with ablation but no collisions (dotted line).}
\label{dustdist}
\end{figure}

Before including the effects of the pulsar wind,
we start with a standard collisional cascade similar to
our own asteroid belt's.
Starting with 10 $\MEarth$ of 500 km radius planetesimals located
at 2.5 AU, after 1 Gyr of collisions the belt of material
has ground down to just $10^{-2} \MEarth$
\citep[c.f. $\sim$10$^{-3} \MEarth$ contained in the asteroid
belt;][]{davis02}.  
The resultant size distribution ({\it dashed line} in Figure~\ref{dustdist})
extends from the large planetesimals down to small dust.
For grain sizes smaller than $\sim$mm, Poynting-Robertson drag begins
to remove dust faster than it can be replenished by collisions,
resulting in a small but significant flattening of the size distribution
for the smallest grains.
In the Solar System, for example, the combination of collisions and
P-R drag result in a typical dust
radius of several 100 \um\ within the asteroid belt \citep{dermott02}
while the interstellar dust particles reaching the Earth have sizes
somewhat smaller \citep[$\sim$100 \um;][]{love93}. 
The total dust area in our collisional model without ablation is 
$5 \times 10^{22}$ cm$^2$.
For a given total mass, varying the unknown material composition tends
to have relatively little effect on this dust area (denser material
has less emitting area per mass, but is more resistant to removal by
radiation pressure; in the simple model considered here these effects
exactly offset each other). 

At the other extreme we also consider the distribution of object 
sizes under the influence of ablation, but with no collisions
({\it dotted line} in Figure~\ref{dustdist}).
This would apply to a very sparse distribution of planetesimals.
As calculated by \citet{miller01}, objects up to several km in size 
are worn down by the pulsar wind over 1 Gyr.
The other important size scale for this distribution is 100 cm,
the stopping distance for the relativistic particles.  
For objects larger than this,
mass loss is independent of size (giving 
equal numbers of objects in linear mass bins or 
$dn/d(log a) \propto a^{3}$) 
while below 100 cm the mass loss is directly proportional to mass 
(giving equal numbers in logarithmic mass bins
or $dn/d(log a) =$ constant). 



When collisions are again considered in addition to the effects 
of ablation ({\it solid line} in Figure~\ref{dustdist}),
objects smaller than km in size are continually
resupplied by the disruptive collisions of larger objects.
This replenishment greatly increases the amount of small dust and rocks
above the simple ablation estimate (compare the {\it dotted} and 
{\it solid} lines in Figure~\ref{dustdist}).
The ability of Poynting-Robertson drag to remove the smallest dust
is unclear.  (P-R drag is caused by particles absorbing photons with
no angular momentum and then re-emitting them in the dust's rotating
frame.  The particles emitted by the pulsar as it spins down, however,
do carry their own positive angular momentum which they would then
impart on the dust grains.)
Regardless of whether or not many sub-micron grains are present, 
their heating by the pulsar wind is very inefficient, as mentioned above.
The larger planetesimals, on the other hand, are thick enough to
absorb the pulsar's emission energy, and should reach temperatures
comparable to the local blackbody temperature (270~K at 2.5~AU).
As in the pure ablation case, the majority of the warm surface area in
the system continues to be contributed by these large $\sim$km-sized objects.
The total emitting cross section for the model distribution
is $4 \times 10^{19}$ cm$^2$.  This is a factor of 5 greater than 
with ablation alone
({\it dotted line}),
but still three orders of magnitude less than
the area around the standard collisional cascade without ablation
({\it dashed line}).
This area at 2.5~AU corresponds to $\ldsd$ of just $2.3 \times 10^{-9}$,
far below our observational limit.

Models in which disks steadily and smoothly grind away with time,
however, fail to describe the variety of debris disks now observed with \spit.
Just considering analytic models like those in Figure~\ref{dustdist}, 
one might conclude that all
systems would grind down their asteroid belts in a similar fashion to
our own, such that no old stars should have observable inner belts.
This is in contrast with systems such as HD 69830, which has a
bright disk of small dust grains orbiting at $\sim$0.5~AU 
\citep[$\ld \sim 2 \times 10^{-4}$;][]{Beichman05irs}.
Despite the old age of this star ($\sim$2 Gyr), the debris has an emitting
surface area more than 1000 times greater than our
asteroid belt's.
As is clear from the collisional models, such a situation cannot
persist for the lifetime of the star, but instead must be a reflection
of some recent spike in activity.
Based on the broad halo of small, short-lived dust grains blown out
from Vega's debris disk, a large recent collisional event is also
inferred in that system \citep{Su05}.
More generally, \citet{rieke05} find that while debris disk frequency
declines as stars age, even old A stars can have significant emission
due to stochastic collisional events, while \citet{bryden06a} 
similarly find many bright disks among even older FGK stars.
Overall, one can conclude that many, if not most, observed debris
disks are observable only because an unusual event has
recently increased their infrared brightness.
Such collisional events will also cause temporary enhancements of  the
infrared emission from debris disks around pulsars.
Although the model in Figure~\ref{dustdist} is four orders of
magnitude below the observational threshold presented in
\S\ref{dustlim}, modest improvements in the detection limit
could detect the aftereffects of a large collisional event like those
commonly seen around main sequence stars,
if one has recently occurred in the \psr\ system.

We conclude that, despite our new limits on IR excess, a 0.01 $\MEarth$ belt
of 100-km-sized asteroids and its collisionally produced dusty debris
may still be present in this system.  
Whether or not this debris could be detected at far-infrared
wavelengths depends strongly on its recent collisional history. 

\acknowledgments {
This publication makes use of NASA/IPAC's InfraRed Science Archive (IRSA)
which provides access to data from the 2MASS and IRAS all-sky surveys.
The {\it Spitzer Space Telescope} is operated by the Jet Propulsion
Laboratory, California Institute of Technology, under NASA contract 1407. 
Development of MIPS was funded by NASA through the Jet Propulsion
Laboratory, subcontract 960785.  Some of the research described in
this publication was carried out at the Jet Propulsion Laboratory,
California Institute of Technology, under a contract with the National
Aeronautics and Space Administration.  We would like to thank 
Tom Kuiper for discussions on pulsar emission and an anonymous
referee for helpful comments.} 


\begin{thebibliography}
\expandafter\ifx\csname natexlab\endcsname\relax\def\natexlab#1{#1}\fi

\bibitem[{{Abazajian} {et~al.}(2005){Abazajian}, {Adelman-McCarthy},
  {Ag{\"u}eros}, {Allam}, {Anderson}, {Anderson}, {Annis}, {Bahcall}, {Baldry},
  {Bastian}, {Berlind}, {Bernardi}, {Blanton}, {Bochanski}, {Yocum}, {York},
  {Zehavi}, {Zibetti}, \& {Zucker}}]{Abazajian05}
{Abazajian}, K., {Adelman-McCarthy}, J.~K., {Ag{\"u}eros}, M.~A., {et~al.}
  2005, \aj, 129, 1755

\bibitem[{{Aumann} {et~al.}(1984){Aumann}, {Beichman}, {Gillett}, {de Jong},
  {Houck}, {Low}, {Neugebauer}, {Walker}, \& {Wesselius}}]{aumann84}
{Aumann}, H.~H., {Beichman}, C.~A., {Gillett}, F.~C., {et~al.} 1984, \apjl,
  278, L23

\bibitem[{{Beckwith} {et~al.}(1990){Beckwith}, {Sargent}, {Chini}, \&
  {Guesten}}]{beckwith90}
{Beckwith}, S.~V.~W., {Sargent}, A.~I., {Chini}, R.~S., \& {Guesten}, R. 1990,
  \aj, 99, 924

\bibitem[{{Begelman} \& {Li}(1992)}]{begelman92}
{Begelman}, M.~C. \& {Li}, Z.-Y. 1992, \apj, 397, 187

\bibitem[{{Beichman} {et~al.}(2005{\natexlab{a}}){Beichman}, {Bryden},
  {Gautier}, {Stapelfeldt}, {Werner}, {Misselt}, {Rieke}, {Stansberry}, \&
  {Trilling}}]{Beichman05irs}
{Beichman}, C.~A., {Bryden}, G., {Gautier}, T.~N., {et~al.} 2005{\natexlab{a}},
  \apj, 626, 1061

\bibitem[{{Beichman} {et~al.}(2005{\natexlab{b}}){Beichman}, {Bryden}, {Rieke},
  {Stansberry}, {Trilling}, {Stapelfeldt}, {Werner}, {Engelbracht}, {Blaylock},
  {Gordon}, {Chen}, {Su}, \& {Hines}}]{beichman05mips}
{Beichman}, C.~A., {Bryden}, G., {Rieke}, G.~H., {et~al.} 2005{\natexlab{b}},
  \apj, 622, 1160

\bibitem[{{Bryden et al.}(2006)}]{bryden06a}
{Bryden et al.}, G. 2006, \apj, 636, 1098

\bibitem[{Cordes \& Lazio(2002)}]{Cordes02}
Cordes, J.~M. \& Lazio, T. J.~W. 2002

\bibitem[{{Cutri} {et~al.}(2003){Cutri}, {Skrutskie}, {van Dyk}, {Beichman},
  {Carpenter}, {Chester}, {Cambresy}, {Evans}, {Fowler}, {Gizis}, {Howard},
  {Huchra}, {Jarrett}, {Kopan}, {Kirkpatrick}, {Light}, {Marsh}, {McCallon},
  {Schneider}, {Stiening}, {Sykes}, {Weinberg}, {Wheaton}, {Wheelock}, \&
  {Zacarias}}]{cutri03}
{Cutri}, R.~M., {Skrutskie}, M.~F., {van Dyk}, S., {et~al.} 2003, VizieR Online
  Data Catalog, 2246

\bibitem[{{Davis} {et~al.}(2002){Davis}, {Durda}, {Marzari}, {Campo Bagatin},
  \& {Gil-Hutton}}]{davis02}
{Davis}, D.~R., {Durda}, D.~D., {Marzari}, F., {Campo Bagatin}, A., \&
  {Gil-Hutton}, R. 2002, Asteroids III, 545

\bibitem[{{Dermott} {et~al.}(2002){Dermott}, {Durda}, {Grogan}, \&
  {Kehoe}}]{dermott02}
{Dermott}, S.~F., {Durda}, D.~D., {Grogan}, K., \& {Kehoe}, T.~J.~J. 2002,
  Asteroids III, 423

\bibitem[{{Dohnanyi}(1968)}]{dohnanyi68}
{Dohnanyi}, J.~S. 1968, in IAU Symp. 33: Physics and Dynamics of Meteors, 486

\bibitem[{{Dole} {et~al.}(2004){Dole}, {Rieke}, {Lagache}, {Puget},
  {Alonso-Herrero}, {Bai}, {Blaylock}, {Egami}, {Engelbracht}, {Gordon},
  {Hines}, {Kelly}, {Le Floc'h}, {Misselt}, {Morrison}, {Muzerolle},
  {Papovich}, {P{\' e}rez-Gonz{\' a}lez}, {Rieke}, {Rigby}, {Neugebauer},
  {Stansberry}, {Su}, {Young}, {Beichman}, \& {Richards}}]{dole04}
{Dole}, H., {Rieke}, G.~H., {Lagache}, G., {et~al.} 2004, \apjs, 154, 93

\bibitem[{{Dominik} \& {Decin}(2003)}]{dominik03}
{Dominik}, C. \& {Decin}, G. 2003, \apj, 598, 626

\bibitem[{{Durda} \& {Dermott}(1997)}]{durda97}
{Durda}, D.~D. \& {Dermott}, S.~F. 1997, Icarus, 130, 140

\bibitem[{{Foster} \& {Fischer}(1996)}]{foster96}
{Foster}, R.~S. \& {Fischer}, J. 1996, \apj, 460, 902

\bibitem[{{Gautier et al.}(2006)}]{gautier06}
{Gautier et al.}, T.~N. 2006, \apj, in preparation

\bibitem[{{Gaensler} \& {Slane}(2006)}]{Gaensler06}
{Gaensler}, B.~M. \& {Slane}, P.~O. 2006, \araa, 44, 1

\bibitem[{{Gordon} {et~al.}(2005){Gordon}, {Rieke}, {Engelbracht}, {Muzerolle},
  {Stansberry}, {Misselt}, {Morrison}, {Cadien}, {Young}, {Dole}, {Kelly},
  {Alonso-Herrero}, {Egami}, {Su}, {Papovich}, {Smith}, {Hines}, {Rieke},
  {Blaylock}, {P{\' e}rez-Gonz{\' a}lez}, {Le Floc'h}, {Hinz}, {Latter},
  {Hesselroth}, {Frayer}, {Noriega-Crespo}, {Masci}, {Padgett}, {Smylie}, \&
  {Haegel}}]{Gordon05a}
{Gordon}, K.~D., {Rieke}, G.~H., {Engelbracht}, C.~W., {et~al.} 2005, \pasp,
  117, 503

\bibitem[{{Gordon et al.}(2005)}]{Gordon05b}
{Gordon et al.}, K.~D. 2005, SPIE, 5487, in press

\bibitem[{{Go{\'z}dziewski} {et~al.}(2005){Go{\'z}dziewski}, {Konacki}, \&
  {Wolszczan}}]{goz05}
{Go{\'z}dziewski}, K., {Konacki}, M., \& {Wolszczan}, A. 2005, \apj, 619, 1084

\bibitem[{{Greaves} \& {Holland}(2000)}]{greaves00}
{Greaves}, J.~S. \& {Holland}, W.~S. 2000, \mnras, 316, L21

\bibitem[{{Gustafson}(1994)}]{gustafson94}
{Gustafson}, B.~A.~S. 1994, Annual Review of Earth and Planetary Sciences, 22,
  553

\bibitem[{{Habing} {et~al.}(2001){Habing}, {Dominik}, {Jourdain de Muizon},
  {Laureijs}, {Kessler}, {Leech}, {Metcalfe}, {Salama}, {Siebenmorgen},
  {Trams}, \& {Bouchet}}]{habing01}
{Habing}, H.~J., {Dominik}, C., {Jourdain de Muizon}, M., {et~al.} 2001, \aap,
  365, 545

\bibitem[{{Kennel} \& {Coroniti}(1984)}]{kennel84}
{Kennel}, C.~F. \& {Coroniti}, F.~V. 1984, \apj, 283, 710

\bibitem[{{Koch-Miramond} {et~al.}(2002){Koch-Miramond}, {Haas}, {Pantin},
  {Podsiadlowski}, {Naylor}, \& {Sauvage}}]{koch02}
{Koch-Miramond}, L., {Haas}, M., {Pantin}, E., {et~al.} 2002, \aap, 387, 233

\bibitem[{{Konacki} \& {Wolszczan}(2003)}]{konacki03a}
{Konacki}, M. \& {Wolszczan}, A. 2003, \apjl, 591, L147

\bibitem[{{L{\" o}hmer} {et~al.}(2004){L{\" o}hmer}, {Wolszczan}, \&
  {Wielebinski}}]{lohmer04}
{L{\" o}hmer}, O., {Wolszczan}, A., \& {Wielebinski}, R. 2004, \aap, 425, 763

\bibitem[{{Lazio} \& {Fischer}(2004)}]{lazio04}
{Lazio}, T.~J.~W. \& {Fischer}, J. 2004, \aj, 128, 842

\bibitem[{{Lin} {et~al.}(1991){Lin}, {Woosley}, \& {Bodenheimer}}]{Lin91}
{Lin}, D.~N.~C., {Woosley}, S.~E., \& {Bodenheimer}, P.~H. 1991, \nat, 353, 827

\bibitem[{{Love} \& {Brownlee}(1993)}]{love93}
{Love}, S.~G. \& {Brownlee}, D.~E. 1993, Science, 262, 550

\bibitem[{{Malov}(2003)}]{malov2003}
{Malov}, I.~F. 2003, Astronomy Letters, 29, 502

\bibitem[{{Marcy} \& {Butler}(1995)}]{marcy95}
{Marcy}, G.~W. \& {Butler}, R.~P. 1995, Bulletin of the American Astronomical
  Society, 27, 1379

\bibitem[{{Mayor} \& {Queloz}(1995)}]{mayor95}
{Mayor}, M. \& {Queloz}, D. 1995, \nat, 378, 355

\bibitem[{{McCaughrean} \& {O'Dell}(1996)}]{mccaughrean96}
{McCaughrean}, M.~J. \& {O'Dell}, C.~R. 1996, \aj, 111, 1977

\bibitem[{{Michel} \& {Dessler}(1985)}]{michel85}
{Michel}, F.~C. \& {Dessler}, A.~J. 1985, Science, 228, 1015

\bibitem[{{Miller} \& {Hamilton}(2001)}]{miller01}
{Miller}, M.~C. \& {Hamilton}, D.~P. 2001, \apj, 550, 863

\bibitem[{{Moshir} \& {et al.}(1990)}]{moshir90}
{Moshir}, M. \& {et al.} 1990, in IRAS Faint Source Catalogue, version 2.0, 1

\bibitem[{{Pavlov et al.}(2006)}]{pavlov06}
{Pavlov et al.} 2006, \apj, in preparation

\bibitem[{{Phillips} \& {Chandler}(1994)}]{phillips94}
{Phillips}, J.~A. \& {Chandler}, C.~J. 1994, \apjl, 420, L83

\bibitem[{{Podsiadlowski} {et~al.}(1991){Podsiadlowski}, {Pringle}, \&
  {Rees}}]{podsiadlowski91}
{Podsiadlowski}, P., {Pringle}, J.~E., \& {Rees}, M.~J. 1991, \nat, 352, 783

\bibitem[{{Rieke} {et~al.}(2005){Rieke}, {Su}, {Stansberry}, {Trilling},
  {Bryden}, {Muzerolle}, {White}, {Gorlova}, {Young}, {Beichman},
  {Stapelfeldt}, \& {Hines}}]{rieke05}
{Rieke}, G.~H., {Su}, K.~Y.~L., {Stansberry}, J.~A., {et~al.} 2005, \apj, 620,
  1010

\bibitem[{{Stern}(1996)}]{Stern96}
{Stern}, S.~A. 1996, \aap, 310, 999

\bibitem[{{Stevens} {et~al.}(1992){Stevens}, {Rees}, \&
  {Podsiadlowski}}]{stevens92}
{Stevens}, I.~R., {Rees}, M.~J., \& {Podsiadlowski}, P. 1992, \mnras, 254, 19P

\bibitem[{{Su et al.}(2005)}]{Su05}
{Su et al.}, K. 2005, \apj, submitted

\bibitem[{{Tavani} \& {Brookshaw}(1992)}]{tavani92}
{Tavani}, M. \& {Brookshaw}, L. 1992, \nat, 356, 320

\bibitem[{{Taylor} \& {Cordes}(1993)}]{taylor93}
{Taylor}, J.~H. \& {Cordes}, J.~M. 1993, \apj, 411, 674

\bibitem[{{van Buren} \& {Terebey}(1993)}]{vanburen93}
{van Buren}, D. \& {Terebey}, S. 1993, in ASP Conf. Ser. 36: Planets Around
  Pulsars, 327--333

\bibitem[{{Wolszczan}(1994)}]{wolszczan94}
{Wolszczan}, A. 1994, Science, 264, 538

\bibitem[{{Wolszczan} \& {Frail}(1992)}]{wolszczan92}
{Wolszczan}, A. \& {Frail}, D.~A. 1992, \nat, 355, 145

\bibitem[{{Wolszczan} {et~al.}(2000){Wolszczan}, {Hoffman}, {Konacki},
  {Anderson}, \& {Xilouris}}]{wolszczan00}
{Wolszczan}, A., {Hoffman}, I.~M., {Konacki}, M., {Anderson}, S.~B., \&
  {Xilouris}, K.~M. 2000, \apjl, 540, L41

\bibitem[{{Wolszczan} \& {Konacki}(2006)}]{wolszczan06}
{Wolszczan}, A. \& {Konacki}, M. 2006
\end{thebibliography}

\clearpage
\begin{figure}
\begin{center}
\includegraphics[angle=-90,width=6.5in]{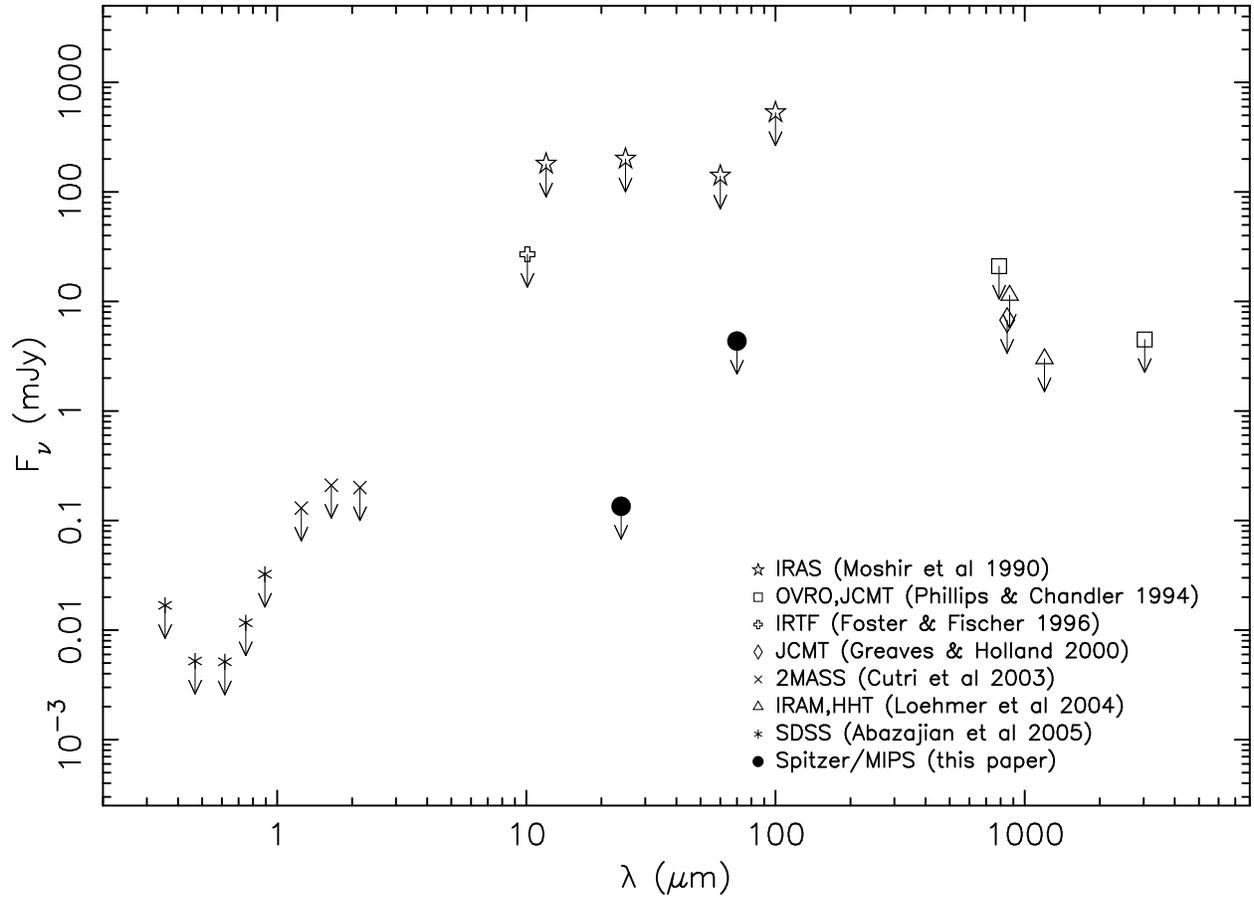} 
\end{center}
\caption{Observational upper limits for emission from the \psr\ system.
Over a range of wavelengths from visible to mm, 
3-$\sigma$ upper limits are shown for data from various sources
(see legend).
Our MIPS 24 and \70um results are shown as filled circles.}
\label{obs}
\end{figure}

\end{document}